\documentclass[9pt,conference]{IEEEtran}

\usepackage{csquotes}
\usepackage{textgreek}
\usepackage{tabularx}
\usepackage{multirow}
\usepackage{multicol}
\usepackage{soul}
\usepackage{graphicx} 
\usepackage{dblfloatfix}    
\usepackage{textcomp}
\usepackage{booktabs} 
\usepackage{graphicx}
\graphicspath{{figs/}}
\usepackage{braket}
\usepackage{multirow}
\usepackage{cite}
\usepackage{amsmath,amssymb,amsfonts}
\usepackage{algorithmic}
\usepackage{graphicx}
\usepackage{textcomp}
\usepackage{xcolor}
\def\BibTeX{{\rm B\kern-.05em{\sc i\kern-.025em b}\kern-.08em
    T\kern-.1667em\lower.7ex\hbox{E}\kern-.125emX}}

\begin{document}

\title{Addressing Temporal Variations in Qubit Quality Metrics for Parameterized Quantum Circuits}

\author{
      Mahabubul Alam, Abdullah Ash-Saki, Swaroop Ghosh \\
  	\begin{tabular}{c}	
     {\normalsize Department of Electrical Engineering} \\
     {\normalsize Pennsylvania State University, University Park, PA-16802} \\
     {\normalsize mxa890@psu.edu, axs1251@psu.edu, szg212@psu.edu} \\
  	\end{tabular}  
  	}

\maketitle

\begin{abstract}
The public access to noisy intermediate-scale quantum ($NISQ$) computers facilitated by  $IBM$, $Rigetti$, $D-Wave$, etc., has propelled the development of quantum applications that may offer quantum supremacy in the future large-scale quantum computers. Parameterized quantum circuits ($PQC$) have emerged as a major driver for the development of quantum routines that potentially improve the circuit's resilience to the noise. $PQC$'s have been applied in both generative (e.g. generative adversarial network) and discriminative (e.g. quantum classifier) tasks in the field of quantum machine learning.
$PQC$'s have been also considered to realize high fidelity quantum gates with the available imperfect native gates of a target quantum hardware. Parameters of a $PQC$ are determined through an iterative training process for a target noisy quantum hardware. \textit{However, temporal variations in qubit quality metrics affect the performance of a $PQC$. Therefore, the circuit that is trained without considering temporal variations exhibits poor fidelity over time.} In this paper, we present training methodologies for $PQC$ in a completely classical environment that can improve the fidelity of the trained $PQC$ on a target $NISQ$ hardware by as much as 42.51\%. 
\end{abstract}

\section{Introduction}\label{intro}

Quantum computing has observed a shift from being a purely academic exploration to a realistic industrial technology in recent years. However, the qubits have small coherence time (i.e., the quantum states are short-lived), the gate operations are imperfect, and the overall computation is extremely error-prone. Moreover, the near-term quantum devices offer a limited number of qubits without the costly feature of error correction. Due to these limitations, it is impossible to implement and test the target quantum algorithms (e.g. shor's factorization, grover's search, etc.) which have made quantum computing so attractive on a useful scale on these noisy intermediate-scale quantum ($NISQ$) hardware. In recent years, quantum routines have been developed which are inherently resilient to errors using variational/parameterized quantum circuits ($PQC$)  
\cite{farhi2014quantum, kandala2017hardware, romero2017quantum}. \textit{$PQC$ is composed of a set of parameterized single and controlled single qubit gates. The parameters are iteratively optimized by a classical optimizer to attain a desired input-output relationship.} For example, RZ($\theta$) gate available in Rigetti 8Q-Agave hardware can be used to perform an arbitrary amount of rotation of a target qubit along Z-axis. By employing variational hybrid quantum/classical algorithms, $PQC$'s have been applied to accomplish both the generative and discriminative tasks in the field of quantum machine learning \cite{dallaire2018quantum, du2018expressive, schuld2018circuit, farhi2018classification, killoran2018continuous, romero2019variational}. For example, Romero et al. proposed a generative variational circuit that consists of two parts: a quantum circuit employed to encode a classical random variable into a quantum state, and a $PQC$ whose parameters are optimized to mimic a target probability distribution \cite{romero2019variational}. Schuld et al. \cite{schuld2018circuit} proposed a low-depth variational quantum algorithm for supervised learning where the input classical feature vectors are encoded into the amplitudes of a quantum system, and a quantum circuit of parameterized single and two-qubit gates together with a single-qubit measurement is used to classify the inputs. In \cite{heya2018variational}, $PQC$'s are used to develop arbitrary high-fidelity quantum gates with the imperfect native gates of a target hardware.

\textbf{Motivation:} The trained $PQC$ is supposed to be noise resilient as the training is generally performed with the noisy hardware in the loop approach to address the impact of noise as shown in Figure \ref{fig:training}(a) \cite{zhu2018training, leyton2019robust, benedetti2018generative}. However, the quantum computers operate under extremely controlled environment (i.e. operating temperature is in millikelvin range \cite{otterbach2017unsupervised}) and the qubit performance metrics that define the qubit quality (e.g. T1 relaxation time, T2 dephasing time, single-qubit gate error, multi-qubit gate error, readout error, etc.) experience significant fluctuations over time. Generally, the quantum computers (e.g., IBMQX4 and IBMQX2 from IBM) are periodically calibrated through randomized benchmarking \cite{knill2008randomized} and the updated qubit quality metrics are reported for the users to validate their quantum experiments on any target hardware. The variations in the performance metrics of the qubits in IBMQX4 quantum computer is shown in Figure \ref{fig:qvar}. The data has been collected over a 43 days period. \textit{The significant variations in the qubit quality metrics indicate that variational circuits that are trained at any particular time using the hardware in the loop training methodology may not show the desired behavior all the time.}

\begin{figure} [!ht] 
\vspace{-1em}
 \begin{center}
    \includegraphics[width=0.5\textwidth]{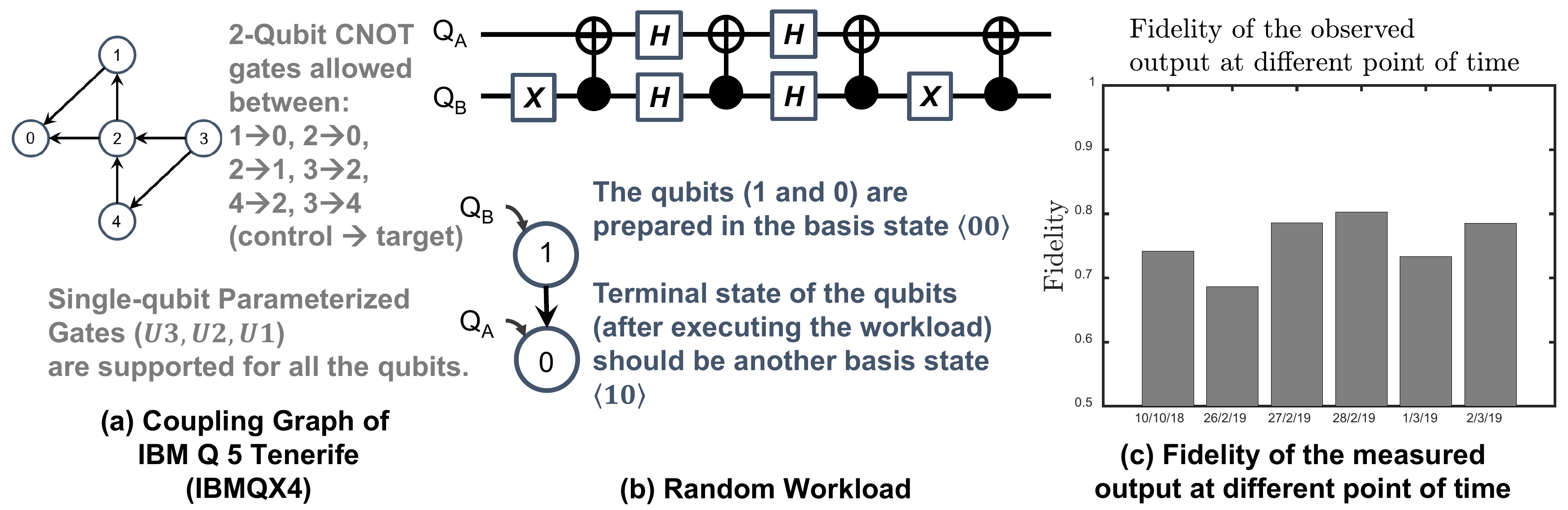}
 \end{center}
 \vspace{-1em}
 \caption{(a) Coupling graph of IBMQX4 (Tenerife) hardware from IBM; (b) Random quantum workload; (c) outcome at different points in time.} \label{fig:mot}
\end{figure}

\begin{figure*} [!ht] 
\vspace{-1em}
 \begin{center}
    \includegraphics[width=0.95\textwidth]{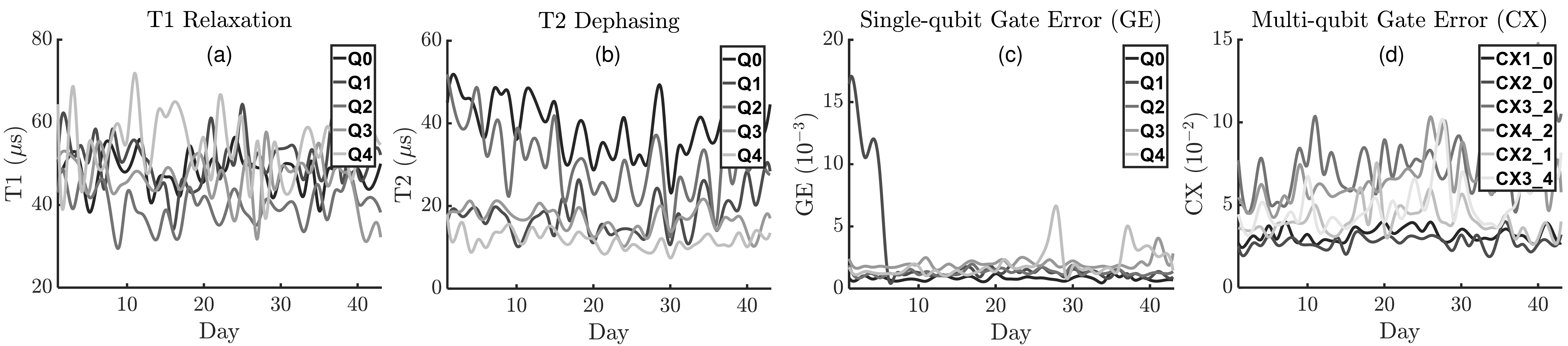}
 \end{center}
 \vspace{-1em}
 \caption{Temporal variations in qubit performance metrics for IBMQX4 (a) T1 relaxation; (b) T2 dephasing; (c) single qubit gate error; (d) two-qubit gate error.} \label{fig:qvar}
 \vspace{-1em}
\end{figure*}

The temporal variability at the output of a quantum circuit is expected for any arbitrary quantum circuit. As a motivational example, we have executed the workload shown in Figure \ref{fig:mot}(b) on 5-qubit IBMQX4 quantum computer (the coupling graph of the device is shown in Figure \ref{fig:mot}(a)) on 5 different occasions. The qubits are prepared in the basis state $\ket{00}$. Ideally, at the end of the execution period, the qubits will be in another basis state $\ket{10}$. A projective measurement on the target hardware is expected to generate a measurement of '10' most of the time. However, due to temporal variations of the qubit quality metrics, we have received significantly different outcomes at different points of time as shown in Figure \ref{fig:mot}(c). The y-axis shows the fidelity of the measurements (which is the \% of the correct output for 1024 samples at a time). For circuits such as circuit-centric binary quantum classifiers based on $PQC$ (discussed in Section \ref{sec:BQC}), the final outcome is decided after analyzing the measurement distributions in a classical computer which can be completely wrong due to the temporal variations of the qubit quality metrics. Moreover, quantum computers are expected to operate in a client-server mode (for reliable hardware operation, the quantum computer is kept in extremely controlled environment). Training of an arbitrary $PQC$ of a client with the target server hardware in the loop approach becomes impractical as the training requires a considerable amount of time and the access to any target hardware through the client-server mode goes through a long wait queue. This is true for IBM and Rigetti that provide free access to their quantum computers 
through a cloud service accessible through qiskit and QCS, respectively. \textit{Hence, fully classical training of $PQC$ to address the variations in qubit quality metrics of a target hardware can be a challenging task.}

\begin{figure} [!hb]
\vspace{-1em}
 \begin{center}
    \includegraphics[width=0.5\textwidth]{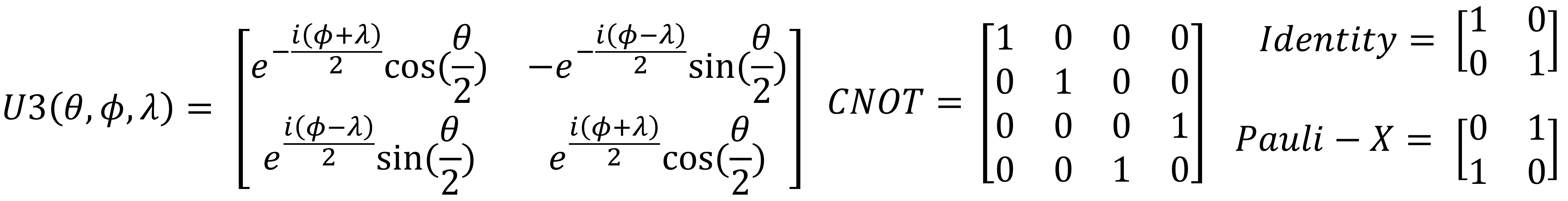}
 \end{center}
 \vspace{-1em}
 \caption{Gate matrices of the quantum gates used in this article.} \label{fig:matrix}
 \vspace{-1em}
\end{figure}

\textbf{Contributions:} In this paper, we, (a) present a framework for simulating any given quantum workload for any target $NISQ$ hardware; (b) demonstrate training methodologies of $PQC$ and address their respective pros and cons; (c) present a fully classical heuristic training methodology for $PQC$ to address the temporal variations in qubit quality metrics; (d) used $PQC$ based circuit-centric quantum classifiers to demonstrate our solutions and verified their effectiveness on real quantum hardware from IBM.

\begin{figure*} [!hb] 
\vspace{-1em}
 \begin{center}
    \includegraphics[width=0.95\textwidth]{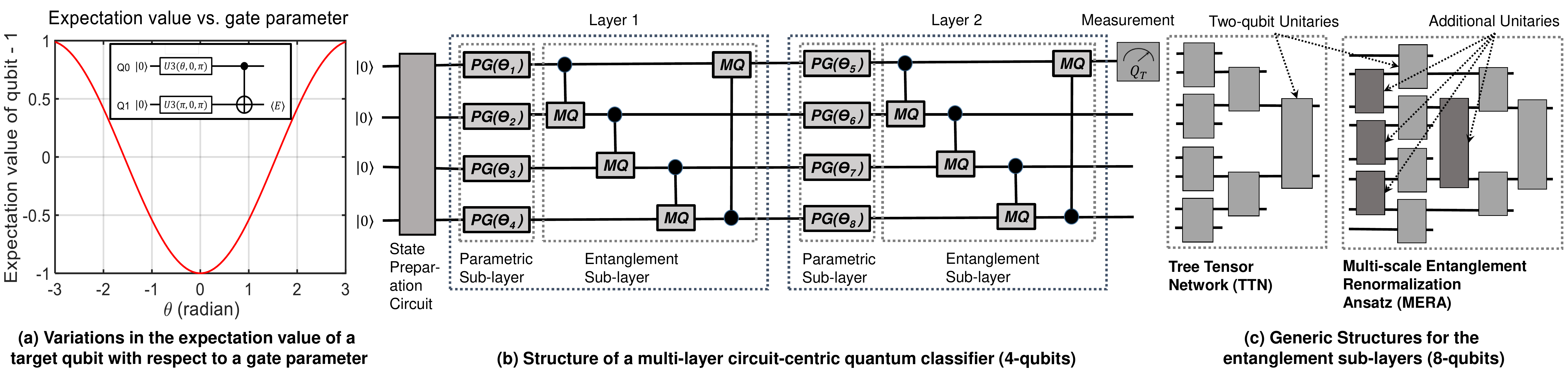}
 \end{center}
 \vspace{-1em}
 \caption{(a) Variations in the expectation value of a target qubit in a variational circuit with respect to the tunable parameter; (b) generic structure of a multi-layer $PQC$ for binary classification task; (b) generic structures for the entanglement sub-layers.} \label{fig:ccqc}
 \vspace{-1em}
\end{figure*}

The paper is organized as follows: the design methodologies of $PQC$ based circuit-centric binary quantum classifier are presented in Section \ref{sec:BQC}. The training methodologies and their pros/cons are is discussed in Section \ref{sec:training}. The framework for modeling circuit behavior on a generic $NISQ$ hardware is discussed in Section \ref{sec:framework}. We demonstrate the proposed training methodology for $PQC$ on two binary quantum classifiers in Section \ref{sec:results}. We conclude in Section \ref{sec:end}.

\section{Binary Quantum Classifiers} \label{sec:BQC}
\subsection{Quantum Computing Preliminaries}
\subsubsection{Qubit and State Vector}
Qubit is the building block of quantum computers. 
Besides storing classical bits 0 and 1, a qubit can be in a superposition of both 0 and 1 simultaneously. Qubit state is expressed with a \textit{ket} ($\ket{.}$) notation which is represented by a column matrix known as \textit{state vector}. A single qubit state $\ket{\psi}$ is described as $\ket{\psi} = a \ket{0} + b \ket{1}$. Here, $\ket{0}$ and $\ket{1}$ are known as computational basis states represented by $[1 \hspace{2mm} 0]^T$ and $[0 \hspace{2mm} 1]^T$ respectively (T stands for matrix transpose), and a and b are complex numbers s.t. $|a|^2 + |b|^2 = 1$. 

\subsubsection{Density Matrix}
An alternate approach of representing qubit state is the \textit{density matrix} ($\rho$) formalism which is expressed as $\rho = \sum_i p_i\ket{\psi}\bra{\psi}$ where $p_i$ is the probability of pure state and $\ket{\psi}$ is the density matrix. This representation is beneficial since qubit states may end up in a mixed state due to noise that can be expressed nicely using density matrix.

\subsubsection{Quantum Gates}
Quantum gates are the operations that modulate the state of qubits and thus perform computations. Mathematically, quantum gates are represented by $2^n \times 2^n$ unitary matrices (n = number of qubits). Quantum gates can work on a single qubit (e.g., Pauli-X ($\sigma_x$) gate) or on multiple qubits (e.g., 2-qubit CNOT gate). When multiple gates work on different qubits, the overall unitary matrix can be calculated using tensor product ($\otimes$). For example, in Fig. \ref{fig:ccqc}(a) two U3 (native gate of IBMQX4) gates are working on qubit-1 and 0. Therefore, the overall gate matrix will be $U = U3 \otimes U3$. The gate matrices of the quantum gates used in this work are shown in Figure \ref{fig:matrix}.

\subsubsection{Expectation Value}
Expectation value is the average of the eigenvalues, weighted by the probabilities that the state is measured to be in the corresponding eigenstate. In quantum computers, measurement of a qubit is performed in the so-called Z-basis or computational basis $\ket{0}$ and $\ket{1}$. These are the eigenvectors (eigenstates) of Pauli-Z ($\sigma_z$) operator with eigenvalues +1 and -1 respectively. For quantum computing, \textit{(a)} a positive expectation value means that the measurements will yield more $\ket{0}$ than $\ket{1}$ , if a qubit prepared in identical setup is measured many times. The measurements will always yield $\ket{0}$ if the expectation value is exactly +1, \textit{(b)} a negative expectation value means that the measurement outcomes will have more $\ket{1}$s than $\ket{0}$s. If the expectation value is exactly -1, the measurements will always yield $\ket{1}$, and \textit{c} if the expectation value is 0, it means the qubit state is in a perfect superposition of both $\ket{0}$ and $\ket{1}$ (e.g., $\psi = (\ket{0} + \ket{1})/\sqrt{2}$) and large number of measurements will result in equal probabilities of $\ket{0}$ and $\ket{1}$.

For more clarity, suppose the state of a qubit after a quantum computation routine is $\ket{\psi} = 0.8\ket{0} + 0.6\ket{1}$ (note the higher amplitude of $\ket{0}$). The expectation value of Pauli-Z operator in this state $\ket{\psi}$ is $\bra{\psi} \sigma_z \ket{\psi}$ = 0.28, a positive expectation value which validates \textit{a} in the above discussion. Figure \ref{fig:ccqc}(a) shows the variations in the expectation value of a target qubit with respect to a gate parameter ($\theta$).

\subsection{Classifier Basics}
Binary classification is the task of classifying any input data into one of two possible groups. In supervised machine learning, this classification problem is solved by training a mathematical model ($f(x,\theta)$) with a properly labeled input data-set \{($x_1$,$y_1$), ($x_2$,$y_2$), .... , ($x_M$,$y_M$)\} where $x_i$ is the feature vector (can be multi-dimensional) of the $i'th$ input data and $y_i$ is the associated label. 
The mathematical model predicts the class of any input data based on its features ($x$) and the parameters ($\theta$) of the model. The parameters ($\theta$) are updated iteratively until the model predictions are satisfactory over the input data-set.

In \cite{schuld2018circuit}, a binary classification on quantum computers is proposed for classical data where a $PQC$ serves as the mathematical model. A state-preparation routine is required to encode the classical data and feed it to the $PQC$. The output is captured from a target qubit. 
During the training phase of the $PQC$, the parameters are updated iteratively based on the given input data-set so that the probability of getting 1 through a measurement of the target qubit for one class is maximized (and 0 for the other class). 

\subsection{State Preparation}
A state preparation circuit (which is applied to the qubits at ground state) is used to convert any classical input data to a quantum format so that quantum gates can be applied on the data and/or quantum speed-up can be exploited. 
The structure of this circuit depends on the chosen encoding scheme. A multitude of quantum encoding scheme of classical data have been proposed \cite{schuld2019quantum}. However, in this paper, we have utilized the basis encoding for parity classification and amplitude encoding for iris classification. These schemes are described below.

\textbf{Basis Encoding:} 
In this scheme, 
binary 0 (1) is encoded as computational basis state $\ket{0}$ ($\ket{1}$). For instance, a classical data x = 9 (binary $1001$) can be represented by 4-qubits (say, $Q_3Q_2Q_1Q_0$) where $Q_3$ and $Q_0$ ($Q_2$ and $Q_1$) are prepared in qubit state $\ket{1}$ ($\ket{0}$).
The effect of the state-preparation routine can be written as - $U_\phi$ : x $\in$ \{ 0, 1 \}$^n$ $\rightarrow$ $\ket{\psi_x}$.

Here, $U_\phi$ is the unitary transformation that prepares the desired quantum state representative of classical data. For IBM quantum computers, all qubits start from a $\ket{0}$ state. Therefore, quantum NOT gate (Pauli-X, $\sigma_x$) has to be applied on $Q_3$ and $Q_0$ whereas Identity gates are applied on $Q_2$ and $Q_1$ to prepare x = 9 state. Thus, for this case $U_{\phi} = \sigma_x \otimes I \otimes I \otimes \sigma_x$. 
Although, the scheme results in a trivial quantum state-preparation circuit (that only requires NOT and Identity gates) which is fairly easy to implement on existing quantum hardware, the required number of qubits may grow linearly with the number of input features (e.g., two 4-bit classical features will require 8 physical qubits). 

\textbf{Amplitude Encoding:} In this scheme, normalized input vectors $x$ = ($x_1$, $x_2$, ....., $x_N$)$^T \in \mathbb{R}$ of dimension $N$ = $2^n$ are associated with the amplitudes of a $n$ qubit state $\ket{\psi_x}$ ($U_\phi$ : x $\in$ $\mathbb{R}$ $\rightarrow$  $\ket{\psi_x}$ = $\sum_{i=1}^{N}$ $x_i$ $\ket{i}$).
\textit{Example:} The state ($\psi$) of a 2-qubit quantum system, due to superposition, is a linear combination of all possible computational basis state i.e. $\psi$ can be written as $a\ket{00} + b\ket{01} + c\ket{10} + d\ket{11}$ such that $\sqrt{a^2 + b^2 + c^2 + d^2} = 1$. Suppose, we have a classical input vector $x = \{1, 2, 3, 4\}$. After normalizing the input vector, we get $x_{norm} = \{0.183, 0.365, 0.547, 0.730\}$. The amplitude encoding scheme will encode this normalized classical input vector entries as the amplitude of the computational basis states of the whole quantum system such that state $\psi$ becomes $0.183\ket{00} + 0.365\ket{01} + 0.547\ket{10} + 0.730\ket{11}$.

In this scheme, the number of qubits grows only logarithmically with the dimension of the classical input vectors (e.g. for the above example, only $log_2(4) = 2$ qubits are required to encode 4 classical values). Furthermore, multiple inputs in superposition state can be processed simultaneously leading to potential speed-up in computation. Mathematically, quantum algorithms that are only polynomial in the number $n$ of qubits can perform computations on the $2^n$ amplitudes leading to a poly-logarithmic processing time.
However, the encoding scheme results in a non-trivial state-preparation circuit which can be unsuitable for existing resource limited quantum hardware.

\subsection{Model Circuit}
The model circuit is a parameterized unitary transformation $U_\theta$ (where $\theta$ is a set of trainable variables) that acts as the mathematical model for the classification task. 
The model circuit  transforms encoded state $\ket{\psi_x}$ to another state, say, $\psi'$ ($\ket{\psi'}$  = $U_\theta$ $\ket{\psi_x}$).
Generally, the model circuit has a layered architecture. Each layer can have identical or dissimilar constructs. A single layer consists of a parametric and an entanglement sub-layer as shown in Figure \ref{fig:ccqc}(b). The parametric sub-layer consists of the parametric single qubit gates ($PG$($\theta$) in Figure \ref{fig:ccqc}(b)). These parameters ($\theta$) are updated during training in an iterative fashion. The entanglement sub-layer consists of multi-qubit gates ($MQ$ gates shown in Figure \ref{fig:ccqc}(b)) which create a dependency between the target qubit and all other qubits in the circuit. The state preparation and model circuit is executed, the state of the target qubit is measured, and these execution and measurement operations are repeated multiple times. The measured distribution is analyzed in a classical computer to determine the class of a single input during inferencing.

\subsection{Design Considerations} 
The selection of gates for the model circuit depends on the available native gates of the target NISQ hardware. For instance, Rigetti 8-Q Agave quantum hardware only supports parametric single-qubit RZ($\theta$) operation with a single rotational parameter ($\theta$). IBMQX4 and IBMQX2 supports parametric single-qubit U3($\theta$, $\phi$, $\lambda$) operation with 3 rotational parameters. The entanglement is realized by applying multi-qubit unitaries on qubits (multi-qubit gates like CNOT or CZ). IBM quantum computers support two-qubit CNOT gates between neighbouring qubits and Rigetti 8-Q Agave supports two-qubit CZ gates. Tree-like structures ($TTN$) have been proposed for the entanglement sub-layer as shown in Figure \ref{fig:ccqc}(c) \cite{huggins2018towards}. $MERA$'s are similar to $TTN$'s, but make use of additional unitary transformations to effectively capture a broader range of quantum correlations as shown in Figure \ref{fig:ccqc}(c) \cite{cincio2008multiscale}. A CNOT gate between two neighbouring qubits in entanglement sub-layer in Rigetti hardware is compiled to 6 unitary transformations resulting in a manifold increase of the depth of the model circuit. Moreover, the CNOT's are allowed in limited directions in hardware such as, IBMQX4 or IBMQX2 which is known as the coupling constraint (e.g., the directional coupling graph of IBMQX4, Figure \ref{fig:mot}(a)). If the model circuit has a CNOT gate that violates the coupling constraint, a swap insertion procedure is executed during the compilation process to ensure that a desired CNOT operation takes place between two target qubits which also increases the depth of the circuit significantly \cite{zulehner2018efficient}. Higher-depth circuits are more susceptible to decoherence induced errors which is the prominent source of error for qubits with a short lifetime. \textit{Therefore, the entanglement sub-layer structure should be chosen based on the available native gates and coupling graph with a goal to minimize the depth of the circuit.}

\begin{figure} [!h] 
\vspace{-1em}
 \begin{center}
    \includegraphics[width=0.45\textwidth]{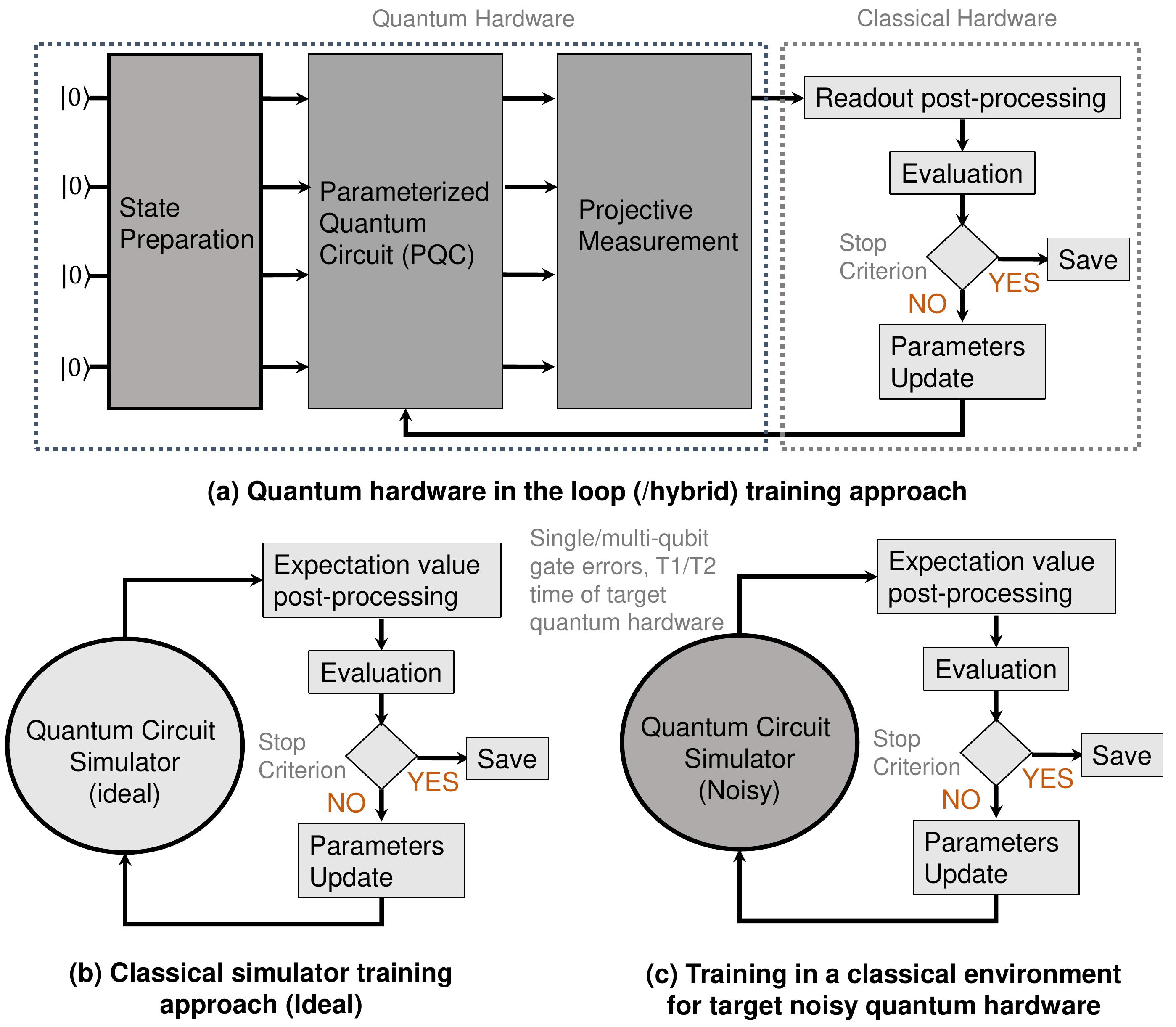}
 \end{center}
 \vspace{-1em}
 \caption{Training of variational circuits ($PQC$) in, (a) quantum-classical hybrid setup; (b) fully classical setup considering an ideal target hardware; (c) fully classical setup for a target noisy hardware.} \label{fig:training}
 \vspace{-1em}
\end{figure}

\section{Training of $PQC$} \label{sec:training}

Training of a model circuit ($PQC$, Figure \ref{fig:ccqc}(b)) for binary classification can follow three disparate strategies as described below. 
\subsection{Existing Approaches}
Two classes of $PQC$ training proposals exist in the literature: 

i) Train the $PQC$ in a hardware-in-the-loop fashion. Hereafter, we term this approach as $app01$. In this approach, the $PQC$ is executed on a real quantum computer. For a certain input, the output is measured and then the measured output is post-processed in a classical computer. Statistical techniques such as, Kullback-Leibler (KL) divergence method is used to calculate the disparity between the target distribution and the measured distribution (hence the cost) to update the parameters with any classical optimization techniques such as stochastic gradient descent or particle swarm optimization etc. \cite{zhu2018training, benedetti2018generative, leyton2019robust}. Then, the $PQC$ is executed again with updated parameters and process iterates until measured output matches target output up to a certain threshold. While it may seem to be an ideal approach, the technique is plagued with certain impediments. \textit{First,} qubits quality changes over time (Fig. \ref{fig:qvar}) which means that a trained $PQC$ on a certain day may not show optimal behavior over time due to qubit specification drift. \textit{Second,} the quantum computers are expected to operate in a client-server fashion. Iterative training scheme may get prohibitively lengthy. Moreover, unlike classical bit states, intermediate quantum-mechanical states cannot be saved in a memory for computation at a later stage since the saved states will be lost due to decoherence. 

ii) Simulation based training of the $PQC$ where a model quantum computer is simulated (we name it $app02$). The simulation results in the expectation value of the result qubit which is then compared with target expectation value to calculate the cost. Now, we can define the following cost-function to iteratively update the parameters of the $PQC$ (Figure \ref{fig:training}(b)) to solve the binary classification problem (described for the hybrid approach) \cite{schuld2018circuit}: 

\begin{equation} \label{eqn:classical}
J(\theta)  = \frac{1}{m} \sum_{i=1}^{m} (y_i - expectation(PQC(x_{i},\theta):Q_T))^2
\end{equation}

where $m$ is the batch-size, $y_i$ is the label of the $i'th$ data in the batch (data are labeled as -1 and +1 for class A and class B respectively), $x_i$ is the $i'th$ input, and 'expectation(PQC($x_{i}$,$\theta$):$Q_T$)' is the expectation value of the target qubit ($Q_T$) for the $i'th$ input and current values of the $\theta$. The target is to minimize the cost. Gradient descent technique is applied to achieve the optimization goal where the partial derivatives of the cost function (Equation \ref{eqn:classical}) with respect to the circuit parameters are calculated using numerical differentiation \cite{schuld2018evaluating}.

In this approach, the client need not wait for the server (quantum hardware) to train and get the parameters of $PQC$. However, the simulation models an ideal (i.e., without noise) quantum computer whereas quantum computers are noisy (and noise behavior shows temporal variation) as pointed out in Section \ref{intro}. Therefore, the parameter optimization without considering noise may not give optimal result during inferencing phase in the real noisy quantum computer.

\subsection{Proposed Approach: Classical Training with Noise Effects}\label{app3}
To deal with the noisy hardware related dependency of the trained $PQC$, we propose to update the parameters where the expectation values are calculated with modeled noise behavior of a target hardware with our noisy quantum hardware simulation framework (described in Section \ref{sec:framework}). The cost function remains same as in Equation \ref{eqn:classical}.
To address the stochastic behavior of the noise sources as evident from Figure \ref{fig:qvar}, we 
use the average value of the qubit quality metrics collected over a significant amount of time (43 days) to optimize the $PQC$ parameters. Before averaging, outliers are removed from the data-set using an interquartile range rule \cite{barbato2011features}. We term this approach as $app03$.
It is expected that circuits optimized with $app03$ will perform better than circuits optimized with $app01$ but executed on a different day and $app02$. In Section \ref{sec:results} will provide sufficient evidence behind this claim, both from simulation and real quantum computer.

\section{Modeling and Simulation Setup} \label{sec:framework}
\subsection{Modeling of Noisy Quantum System}
\subsubsection{Gate Error}\label{subsubsec:gate_erro}
To simulate 1-qubit and 2-qubit gate errors, depolarizing noise channel is applied. Under depolarizing noise, the qubit retains its state with a probability of $(1-p)$ (p = probability of error) and undergoes X (bit-flip), Z (phase-flip) and Z (bit-phase-flip) errors with a probability of $(p/3)$ each.

\subsubsection{T1 Relaxation}
T1 relaxation is simulated with amplitude damping channel. 
Note that the T1 relaxation affects only the state $\ket{1}$ (i.e., $\ket{1} \rightarrow \ket{0}$) leaving state $\ket{0}$ invariant. Reported T1 times are converted to probability (of \textit{no} error) using the formula $p = exp(-t/T1)$ where t is the time of operation that depends on the gate-time of a particular quantum computing hardware. For example, for IBMQX4 quantum computer 1-qubit U2 gate-time is about $60ns$ \cite{IBMQ}.  

\subsubsection{T2 Dephasing}
T2 dephasing is simulated with phase damping channel. Phase damping is a quantum-mechanical phenomenon and therefore difficult to comprehend intuitively. Mathematically, the off-diagonal elements of a density matrix (representing the qubit state) decay to 0 due to T2 dephasing or phase damping. For example, Bell state ($(\ket{00}+\ket{11})/\sqrt{2}$) is an example of entangled state. If the qubits in Bell state undergo dephasing then eventually the off-diagonal terms in the density matrix become zero and entangled qubits end up in a mixed state. Entanglement is believed to be one of the key properties that fuel quantum computers' computing ability and dephasing is detrimental to that. 
Reported T2 times are converted to probability (of \textit{no} error) using the formula $p = exp(-t/T2)$ where t is the time of operation.

\subsubsection{Operator-sum Representation}
To simulate the effect of noise on the quantum computation, we adopt the operator-sum representation using the appropriate Kraus operators \cite{kraus}. In this representation, a quantum operation $\mathcal{E}()$, that maps the input state $\rho_{in}$ (in density matrix format) to output state $\rho_{out}$ such that $\rho_{in} \mapsto \mathcal{E}(\rho_{in}) = \rho_{out} = \sum_k E_k \rho_{in} E_k^{\dagger}$. $E_k$ is called the operation element. By choosing appropriate operation elements the operator-sum representation can be used to compute the output after applying a gate on a qubit. 
\begin{figure}[tb]
\includegraphics[width=3in]{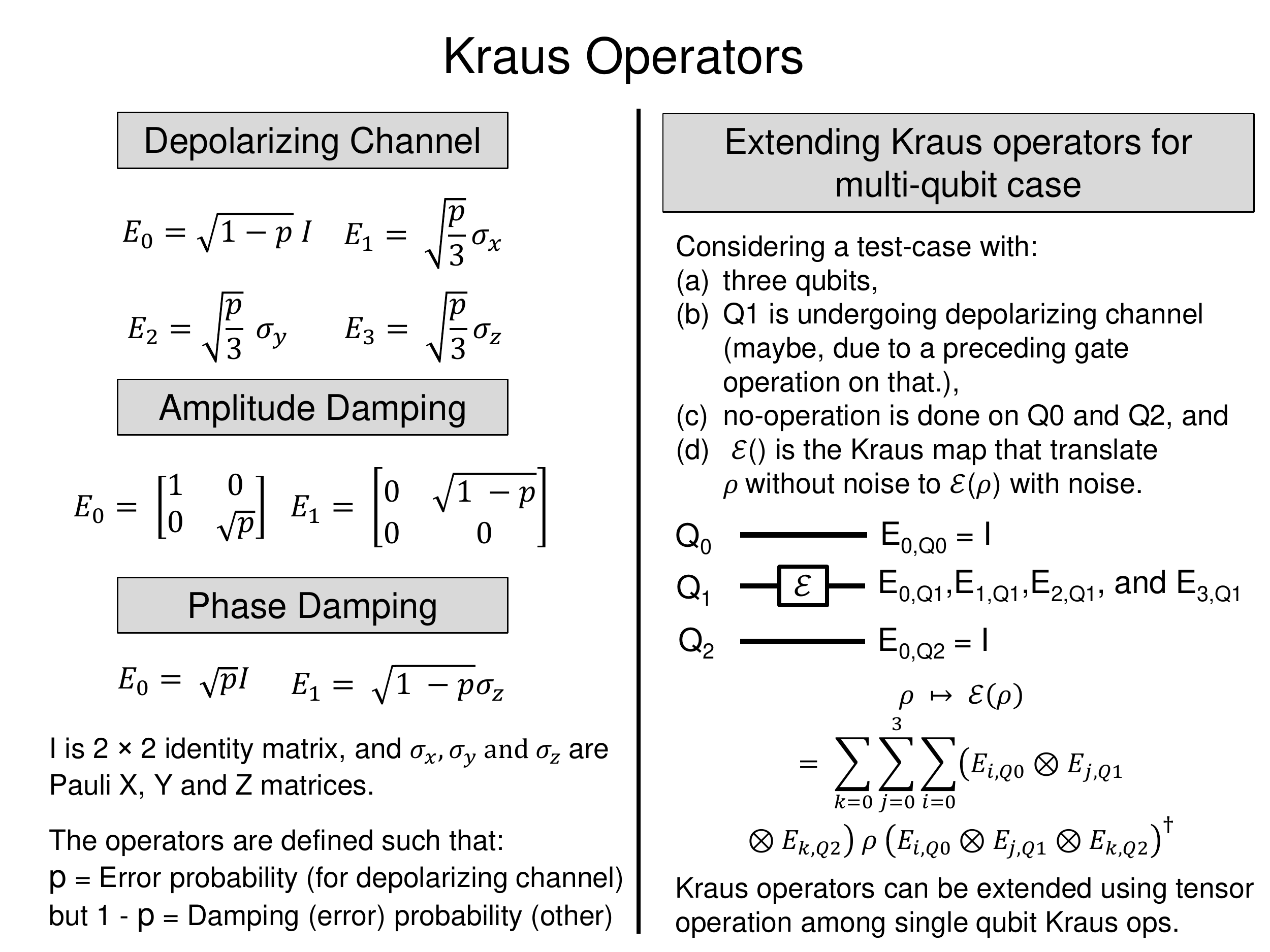}
\vspace{-1em}
\caption{Kraus operators for different noise channels and extending Kraus operators for multi-qubit case. \vspace{-5mm}}
\label{kraus_op}

\end{figure}

Likewise, if appropriate Kraus operators are chosen as operation elements, the operator-sum representation can be used to simulate the effect of different errors on a qubit state. In this paper, we emulate the noisy quantum-processing-units behavior with gate-error and T1 relaxation and T2 dephasing with suitable Kraus operators (listed in Figure \ref{kraus_op}).

\subsubsection{Simulation Flow}
\begin{figure}[!t]
\centering
\includegraphics[width=2.8in]{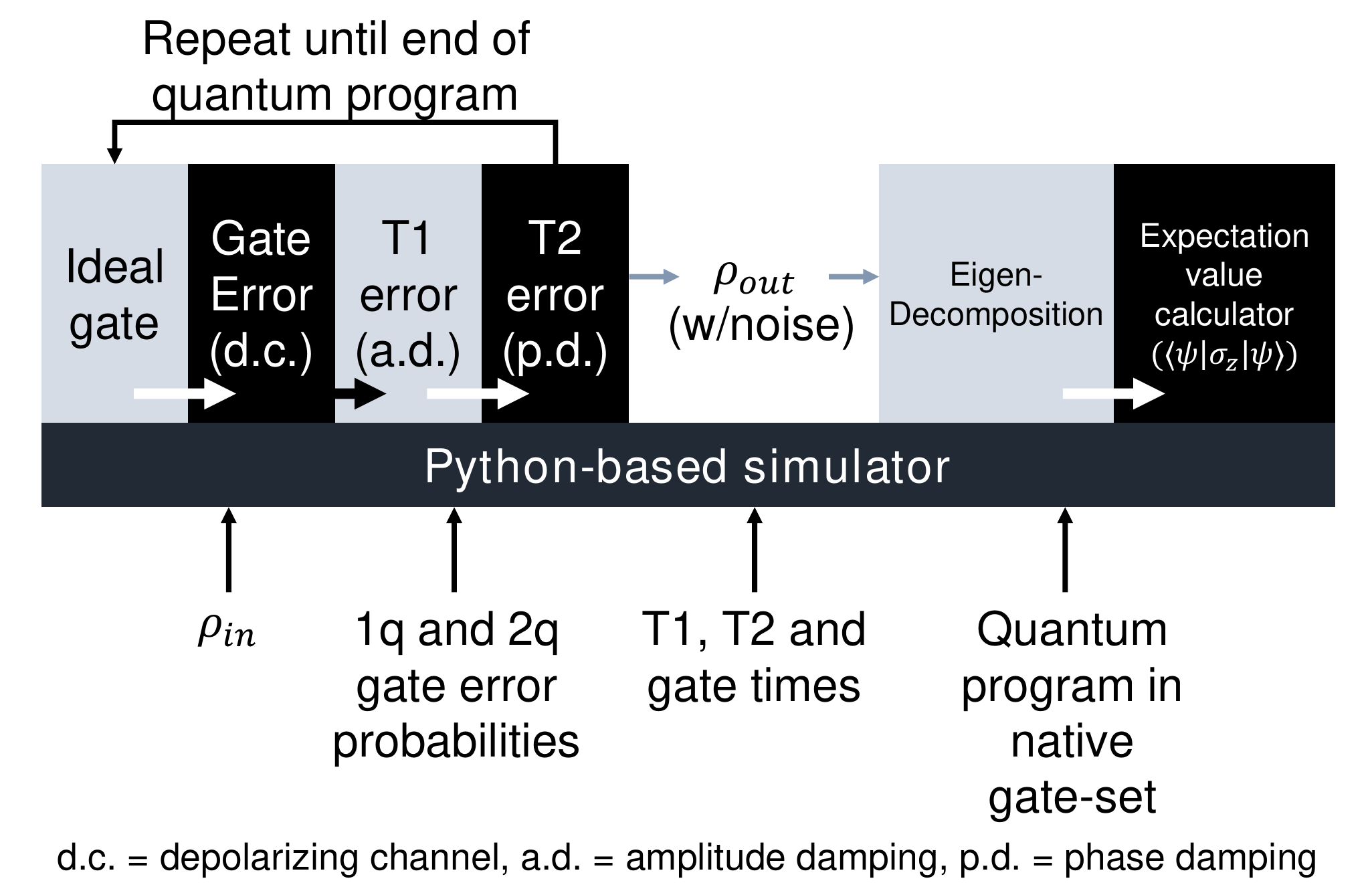}
\vspace{-1em}
\caption{(a) Diagram of the simulation platform and the program flow; (b) 8Q  and modeled 9Q-square architecture. \vspace{-4mm}}
\label{simulator}
\end{figure}

Figure \ref{simulator}(a) shows the schematic of the Python-based simulation platform. We use modules from Qutip \cite{qutip} package to execute matrix operations pertinent to quantum computation. To simulate the behavior of a real quantum device, the simulator takes: (i) the input states of the qubits in density matrix format, (ii) 1-qubit gate error probabilities for each qubit, (iii) 2-qubit gate error probabilities for each allowed qubit pair, (iv) T1 relaxation times, (v) T2 dephasing times, (vi) 1-qubit and 2-qubit gate times and, (vii) a quantum program compiled with native gate-sets of a specific hardware. We primarily used IBMQX4 quantum processing units reported specifications \cite{IBMQ}. However, it can simulate another quantum processor (e.g., Rigetti ASPEN) if appropriate items (ii) - (vii) are fed. 

The simulator reads the quantum program and executes each gate instruction. First, an ideal gate is applied to the qubit states followed by the errors in the sequence gate error, T1 and T2 error. The real quantum device (e.g., IBMQX4) reports amplitudes of pure states as the output. However, the simulator outputs a density matrix ($\rho_{out}$) which contains the result in a possible mixed state. Therefore, the output density matrix is then eigen-decomposed with pure state vectors as eigenvectors. The resulting eigenvalues are the amplitudes of each pure state. For example, if you consider a 2-qubit system, it has 4 possible pure state vectors ($\psi$) i.e. $\ket{00}$, $\ket{01}$, $\ket{10}$ and $\ket{11}$ (each vector is $4 \times 1$). If $\lambda$ is the eigenvalue, then solving $\rho_{out}.\ket{\psi} = \lambda .\ket{\psi}$ will give $\lambda$ and this operation is the eigen-decomposition. The operation has to be repeated for all the pure state vectors (4 in this case) to get all the corresponding eigenvalues (pure state amplitudes). Finally, decomposed state vectors are fed into the expectation value calculator to calculate the expectation value of a qubit ($\langle E \rangle = \bra{\psi}\sigma_z \ket{\psi}$).

\begin{figure}[t]
\centering
\includegraphics[width=2.3in]{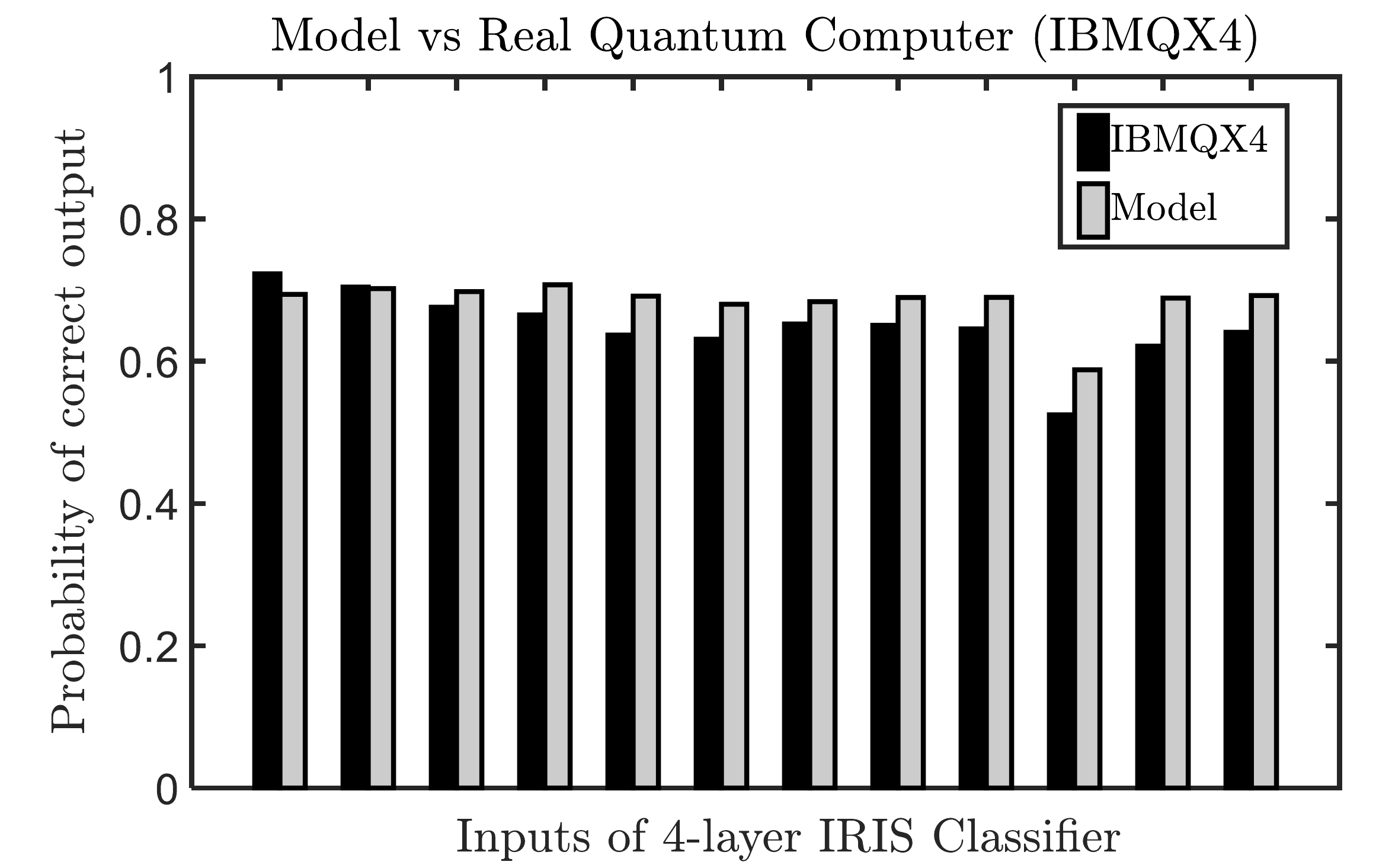}
\vspace{-1em}
\caption{Comparison between fidelities from the simulation model and IBMQX4 real device. \vspace{-2mm} } 
\label{calibration}
\vspace{-1em}
\end{figure}

To validate the model, we simulated the 4-layer IRIS classifier (Fig. 13) as the test circuit using our model with IBMQX4 \cite{IBMQ} specs and the program compiled in IBM native gate-set. The same circuit was executed on IBMQX4 on the same day to get real device results. The comparison between model data and real-device data is shown in Figure \ref{calibration} (IBMQX4 probability of correct output = correct trials/total trials) for 12 different inputs. The model exhibits an average error of about $\approx 7.2\%$. 

\subsection{Validation Setup}
\subsubsection{Data Sources}
In order to validate the effectiveness of our proposed training methodology, we have picked, i) 4-bit parity classification problem (which can be also thought of as a high-fidelity 4-qubit parity gate realization problem using $PQC$ \cite{heya2018variational}) with 16 known inputs/outputs combinations with two output classes (even and odd parity), and, ii) iris classification which is probably the best-known database in pattern recognition literature. The iris data-set contains three different classes (Setosa, Versicolour, and Virginica) of 50 samples per class. Each sample has four distinct features. To convert the iris classification problem into a binary classification task, we have selected 100 samples from Setosa and Versicolour classes. 
\subsubsection{Evaluation Method}
Although parameterized quantum circuits can minimize the effects of noise, it cannot suppress it altogether. 
Therefore, the expectation values cannot be optimized to exactly -1/+1 values for all the inputs during the $PQC$ training period which indicates that a measurement is not guaranteed to result in the desired class output (0/1) for a certain input. Thus, the same circuit is executed multiple times (known as \textit{shots} in IBMQX) and the target qubit is measured in each trial to get a distribution or ratio of 1s to 0s in the output. For binary classification, a large ratio e.g., $>$1 ($<$1) indicates the input belongs to the class represented by logic '1' ('0'). \textit{Example:} A trained parity classifier is executed 1024 times on 4-qubits ($Q_3Q_2Q_1Q_0$) of IBMQX4 with input state $Q_3Q_2Q_1Q_0 = 0100$ (note the input has odd number of 1s i.e. odd-parity) with $Q_0$ being the result/target qubit. The execution resulted in a distribution of `000\textbf{0}': 762 times and `000\textbf{1}' 262 times. The ratio of 1s to 0s of the target qubit is 0.34 ($<1$) which indicates class belongs to logic '0' or odd parity (alternately, correct output `0'/incorrect output `1' = 2.9 $>1$). In an ideal noise-less quantum computer, this ratio of 1s to 0s would have been 0. However, a class decision cannot be taken with confidence when the ratio is close to 1.  In a series of measurements, the goal is to get a high ratio value between the correct and the incorrect outputs from a noisy device. \textit{The ratio between the correct and incorrect outputs is also a representation of the fidelity of the circuit.}

It is to be noted, all the circuit examples presented in this paper has two phases: (i) parameter optimization phase or the training phase and (ii) execution phase or the result phase. We propose a heuristic approach for parameter optimization or training and show the effectiveness of our method by getting more optimal results compared to existing approaches (Section \ref{sec:training}-A) in the execution phase. 

\section{Results and Discussions} \label{sec:results}

\subsection{Test-circuit: Parity Classifier}

\begin{figure} [!b] 
\vspace{-1em}
 \begin{center}
    \includegraphics[width=0.45\textwidth]{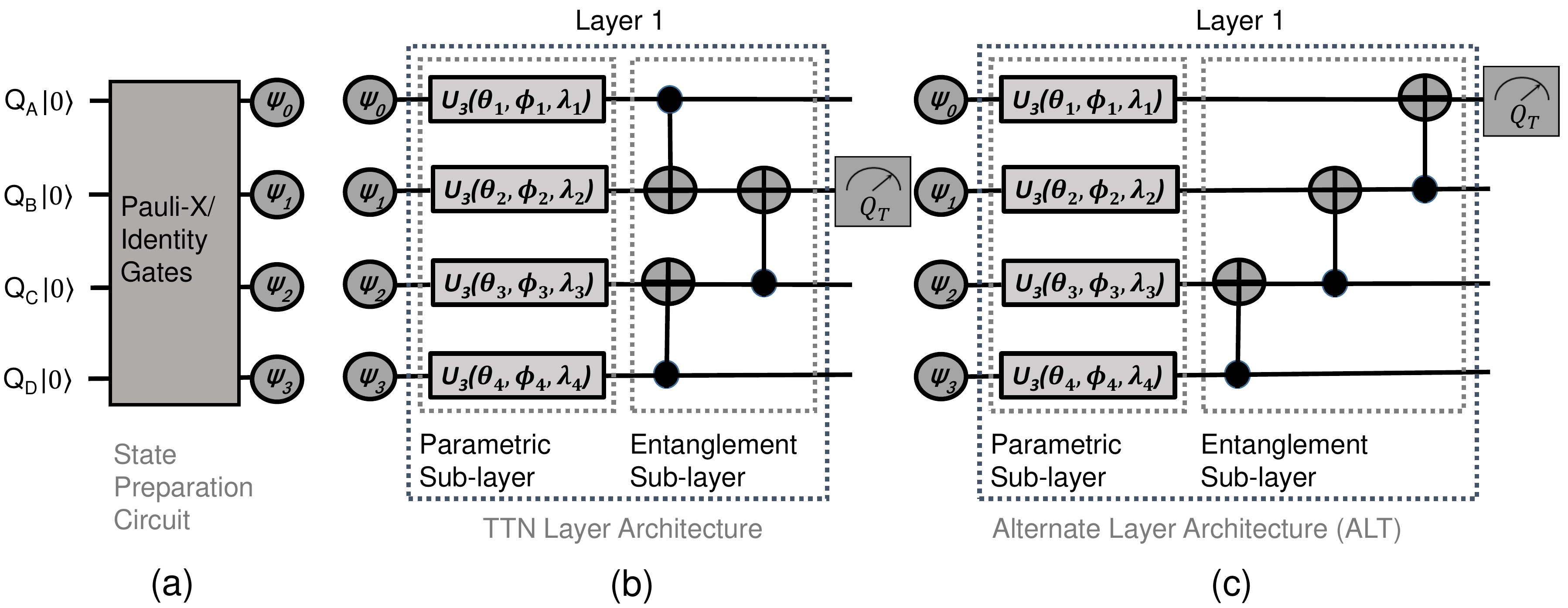}
 \end{center}
 \vspace{-1em}
 \caption{Parity classifier: (a) state preparation circuit, (b) single-layer TTN structure, (c) single-layer ALT structure.} \label{fig:parity_ckt}
 \vspace{-1em}
 \label{fig:9}
\end{figure}

The 4-bit binary inputs for the parity classification is encoded to four qubits using the basis encoding scheme (Section \ref{sec:BQC}). 
Parametric $U3(\theta, \phi, \lambda)$ gates of IBMQX4 have been used as the parametric gates of the model circuits. The `0' outcomes (odd parity), and `1' outcomes (even parity) have been labeled as +1 and -1 respectively for training based on Equation \ref{eqn:classical} using stochastic gradient descent ($SGD$) technique.

\begin{figure} [h] 
\vspace{-1em}
  \begin{center}
     \includegraphics[width=0.45\textwidth]{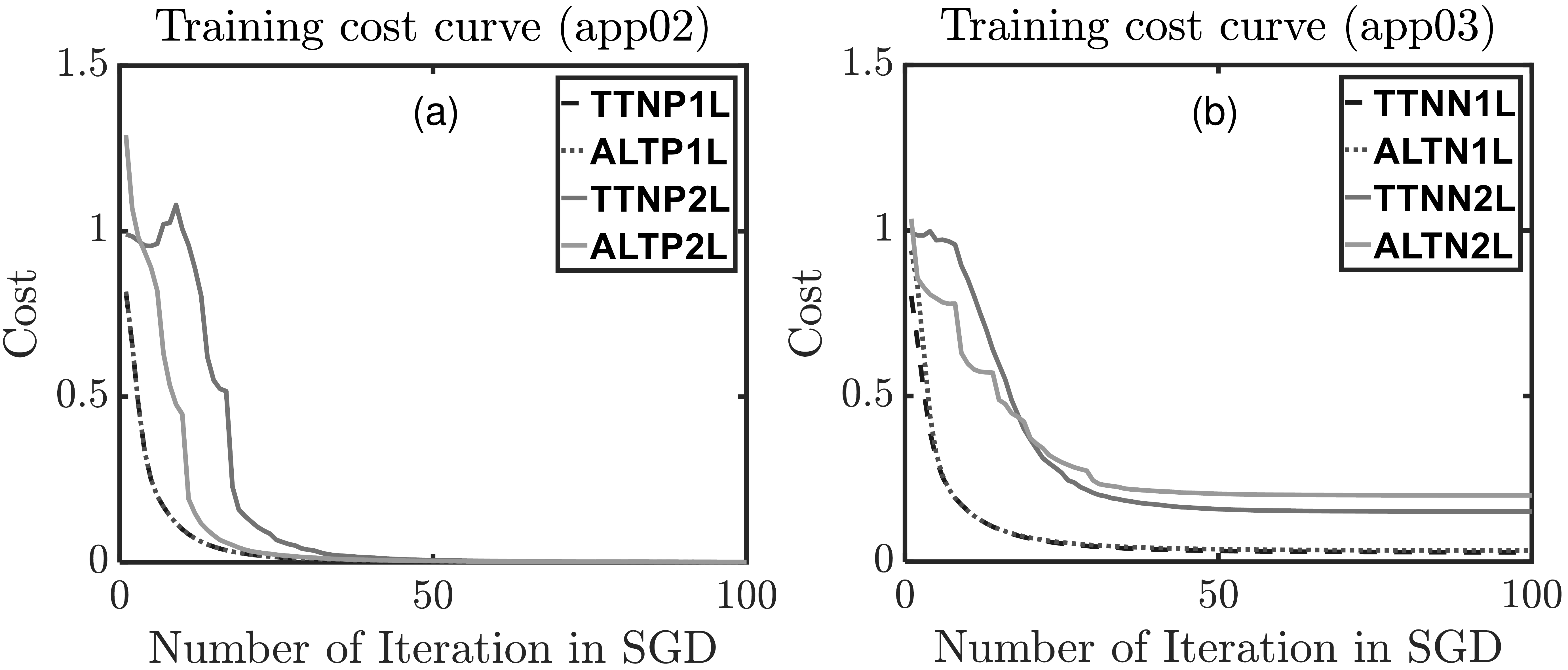}
  \end{center}
  \vspace{-1em}
  \caption{Training cost curves over entire parity classifier input data-set for (a) $app02$ (TTNP1L,TTNP2L,ALTP1L,ALTP2L), (b) $app03$ (TTNN1L,TTNN2L,ALTN1L,ALTN2L),.} \label{fig:cost}
\vspace{-1em}
\end{figure}

We have performed the parameter optimization of the model circuits using three different strategies, $app01$, $app02$ and $app03$ described in Section \ref{sec:training} and tested the performance of the optimized circuits in both real quantum computer, IBMQX4 and in simulation. The cost (m = 16 in Equation \ref{eqn:classical}) over the entire input data-set during the training period ($app02$ and $app03$) is shown in Figure \ref{fig:cost}(a)\&(b). TTNP2L and TTNN2L stand for the cost of two-layer TTN architectures during training for target ideal(\textbf{p}ure) and \textbf{n}oisy hardware respectively. However, it is to be noted that 
we adopted a \textit{simulated} hardware-in-the-loop approach to mimic $app01$ due to interrupted and limited access to the real quantum computer. In this simulated approach, we substitute the $NISQ$ computer with our noisy quantum computer simulation framework (described in Section \ref{sec:framework}) and use the error specifications of a respective day to optimize the $PQC$ parameters.

Note that we have used only 100 iterations of $SGD$ for all the circuits over all the training approaches. The terminal cost, in the parameter optimization or training phase, after 100 iterations are smaller for $app02$ (as evident from Figure \ref{fig:cost}). However, it does not indicate that the trained $PQC$'s would perform better than the ones from $app03$ in the execution phase as the noise characteristics of the real hardware has not been taken into consideration during the optimization procedure for $app02$. In following paragraphs, we show that on a real device $app03$ will always outperform $app02$ which is  substantiated with experiments on a real quantum computer.

\begin{figure} []
 \begin{center}
    \includegraphics[width=0.45\textwidth]{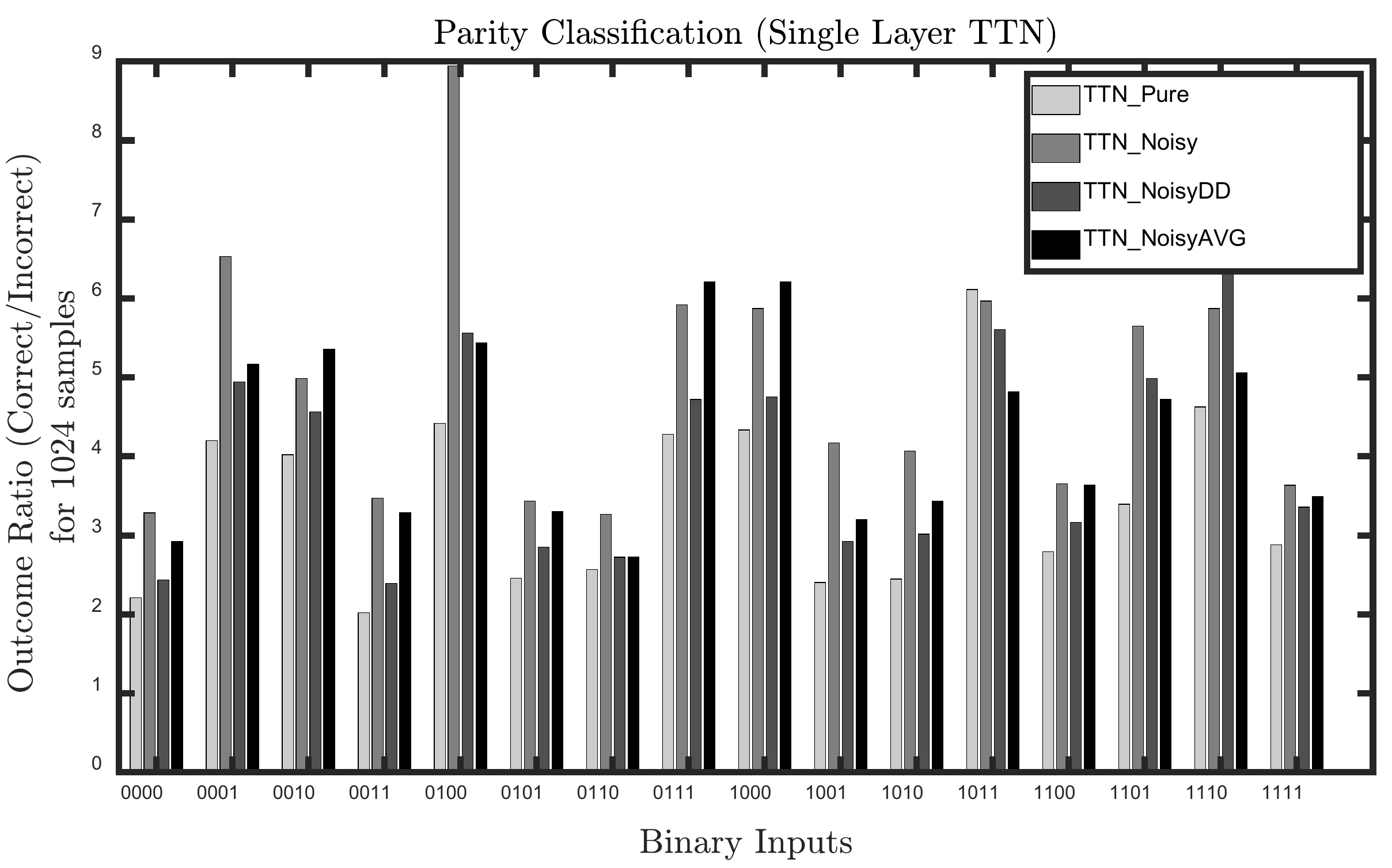}
 \end{center}
 \vspace{-1em}
 \caption{Ratio of correct and incorrect outputs from 1024 samples for different inputs for different training approaches collected from IBMQX4 hardware (app01$\rightarrow$TTN\_Noisy, app02$\rightarrow$TTN\_Pure, app03$\rightarrow$TTN\_NoisyAVG, TTN\_NoisyDD is the optimized $PQC$ in app01 executed on the target hardware in a different day).} \label{fig:parityIBM}
 \vspace{-1em}
 \label{fig:10} 
\end{figure}

Fig. \ref{fig:10} reports the performance (ratio of correct to incorrect outputs) of a binary classifier circuit with 1-layer TTN topology (Fig. \ref{fig:9}) on IBMQX4.  
The $PQC$ generated from the $app01$ approach performed best in terms of the ratio of the correct and incorrect outcome over 1024 repeated measurements on the given day (the day on which the parameters were optimized) as evident from Figure \ref{fig:parityIBM} (TTN\_Noisy). The average of the ratios (TTN\_Noisy) was found 4.92. However, when the same circuit is executed on a different day ((TTN\_NoisyDD in Figure \ref{fig:parityIBM})), it shows random behavior with substantially degraded performance in some cases (average of the ratios: 4.02). This trend validates one of our argument against $app01$ stated in Section \ref{sec:training} i.e. \textit{parameters optimized at one time may not be optimal at a different time.}

The optimized circuit for the $app02$ approach performed poorly (average of the ratios: 3.45) over the entire input data-set (TTN\_Pure in Figure \ref{fig:parityIBM}). 
From the figure, it is evident that the circuit optimized with $app03$ consistently gives better performance than TTN\_Pure and TTN\_NoisyDD corroborating our claim in Section \ref{sec:training}. The ratio of the correct and incorrect outcome for all possible inputs are significantly higher (average of the ratios: 4.31) for the $app03$ approach. It is to be noted TTN\_Pure, TTN\_NoisyDD and TTN\_NoisyAvg data are collected on the same day from IBMQX4.

We further substantiate our claim through simulation with the real hardware being substituted with our NISQ computer simulator in Section \ref{sec:framework}. Two topologies of the parity classifier model circuits ($TTN$ and $ALT$) have been chosen as shown in Figure \ref{fig:parity_ckt}(b)\&(c) both of which satisfies the coupling graph of the IBMQX4 hardware shown in Figure \ref{fig:mot}(a). For each topology, both single-layer and double-layer flavor is simulated i.e. a total of 4 test circuits are simulated. The circuits are optimized with $app02$ (TTNP1L, TTNP2L, ALTP1L, and ALTP2L) and $app03$ (TTNN1L, TTNN2L, ALTN1L, and ALTN2L) to show the superiority of the proposed approach $app03$. 

\begin{figure} [!h] 
 \begin{center}
    \includegraphics[width=0.45\textwidth]{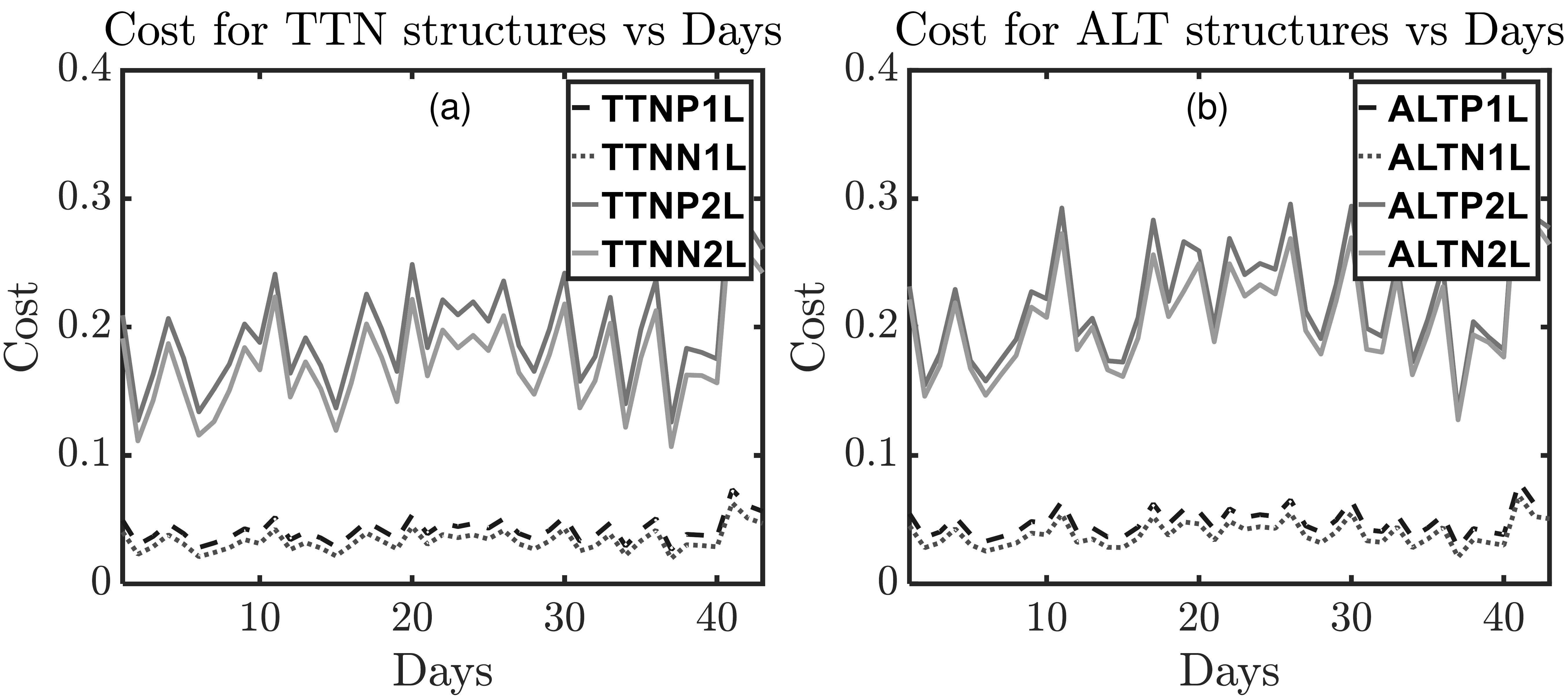}
 \end{center}
 \vspace{-1em}
 \caption{Cost over the entire parity classification data-set over 43 different set of values of the qubit quality metrics of IBMQX4 for, (a) TTN structures; (b) ALT structures. } \label{fig:parity43}
 \vspace{-1em}
\end{figure}

Figure \ref{fig:parity43} shows the aggregate actual cost over the entire input data-set for the trained $PQC$'s ($app02$ - TTNP1L, TTNP2L, ALTP1L, ALTP2L and $app03$ - TTNN1L, TTNN2L, ALTN1L, ALTN2L) for a set of qubit quality metrics data (error specification) of IBMQX4 collected over a 43 days period. The cost here can be interpreted as a measure of the difference between the ideal (expected) result and the result with noise. The lesser the cost the closer the result is to expected. 
The actual cost for the $app03$ is consistently smaller than the $app02$. For instance, the average cost over the entire input data-set for the 43 days period for TTNP1L ($app02$) is 0.042 which is 23.53\% larger than the average cost (0.034) for TTNN1L($app03$).

Thus, both real computer experiments and simulations support our proposed $PQC$ parameter optimization technique considering the noise in real $NISQ$ devices. 
We later executed the optimized $PQC$'s for $app02$ (TTNP1L,TTNP2L) and $app03$ (TTNN1L,TTNN2L) on IBMQX4 for 100 different times with randomly chosen inputs (1024 shots per time) and the cumulative probability ($CP$) distribution of the ratio's of the correct/incorrect outputs ($r$) are shown in Figure \ref{fig:cdf}. The higher ratio values for any given cumulative probability for $app03$ (e.g. TTNN1L $\rightarrow$ $r$ = 5.24 for $CP$ = 0.6) than $app02$ (e.g. TTNP1L $\rightarrow$ $r$ = 4.36 for $CP$ = 0.6) in Figure \ref{fig:cdf} substantiate our previous claim that the $app03$ optimized $PQC's$ would outperform $app02$ on a target quantum hardware. The average value of $r$ for all the $app03$ $PQC$'s were found to be 21.91\% higher than $app02$ for the TTN parity classifiers.

\subsection{Test-circuit: Iris Classifier}
\begin{figure} [t] 
\vspace{-1em}
 \begin{center}
    \includegraphics[width=0.45\textwidth]{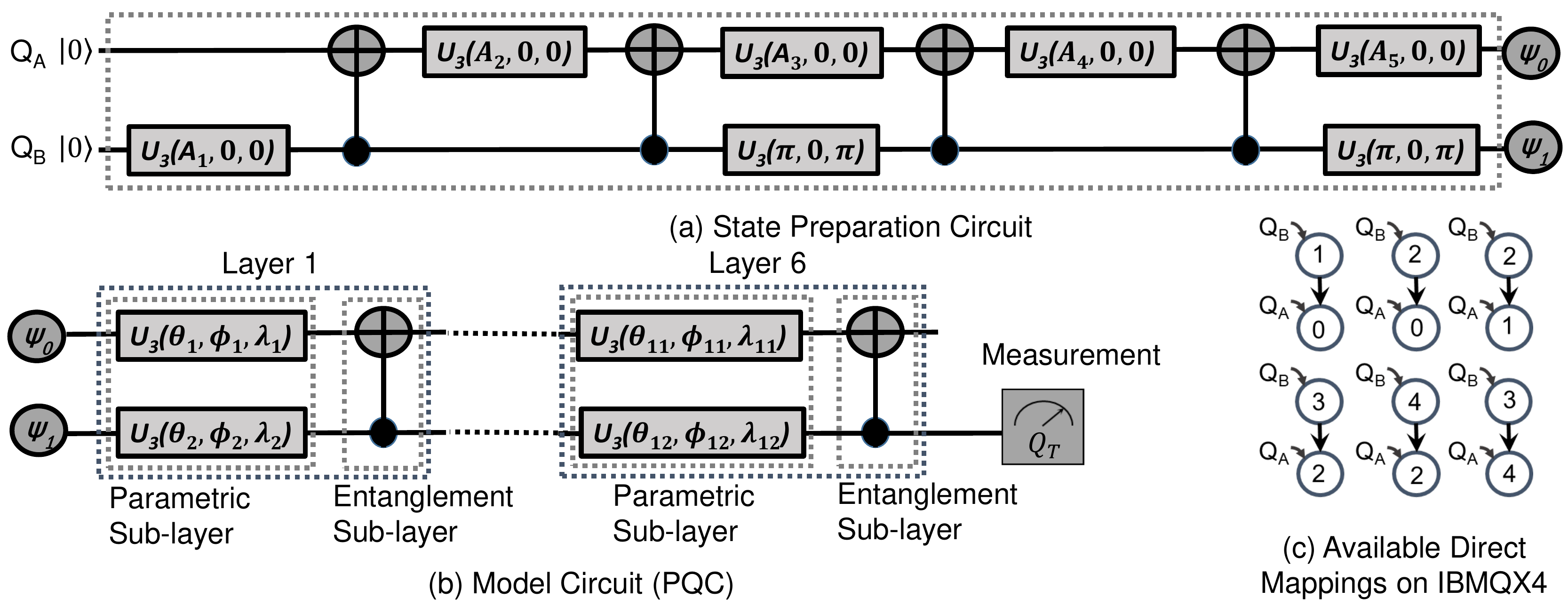}
 \end{center}
 \vspace{-1em}
 \caption{Iris classifier: (a) state preparation circuit, (b) the model circuit ($PQC$), (c) available direct mappings on the target IBMQX4 hardware.} \label{fig:iris_ckt}
 \vspace{-1em}
\end{figure}

We prove our proposal with few more test circuits from IRIS classifier category (Fig. \ref{fig:iris_ckt}). The state preparation circuit has been coded according to the amplitude encoding scheme presented in \cite{mottonen2004transformation}. We have decomposed the controlled Y-axis rotations to native gates available on IBMQX4 \cite{nielsen2002quantum}. The four classical features in a single iris sample are normalized (x[0:3]) as in Section II-C and then used to compute the angles ($A_1$,...,$A_5$) of the state preparation circuit shown in Figure \ref{fig:iris_ckt}(a).

\begin{equation} \label{eqn:iris}
\begin{split}
\beta_0 = 2\arcsin{\frac{x[1]^2}{\sqrt[]{x[0]^2+x[1]^2}}}, \beta_1 = 2\arcsin{\frac{x[3]^2}{\sqrt[]{x[2]^2+x[3]^2}}} \\
\beta_2 = 2\arcsin{\frac{\sqrt[]{x[2]^2+x[3]^2}}{\sqrt[]{x[0]^2+x[1]^2+x[2]^2+x[3]^2}}} \\
A1 = \beta_2, A2 = -A3 = -\beta_{1}/2, A4 = -A5 = -\beta_{0}/2
\end{split}
\end{equation}

The model circuit is composed of multiple layers of parametric $U3(\theta,\phi,\lambda)$ gates and CNOT gates shown in Figure \ref{fig:iris_ckt}(b). We have used two different flavors of the model circuit (4 layers and 6 layers) for validation. The circuit has 6 direct mappings available on IBMQX4 which is shown in Figure \ref{fig:iris_ckt}(c). The samples in the Setosa and Versicolour classes are labeled +1 and -1 respectively for training the model circuit. 

\begin{figure} [] 
 \begin{center}
    \includegraphics[width=0.45\textwidth]{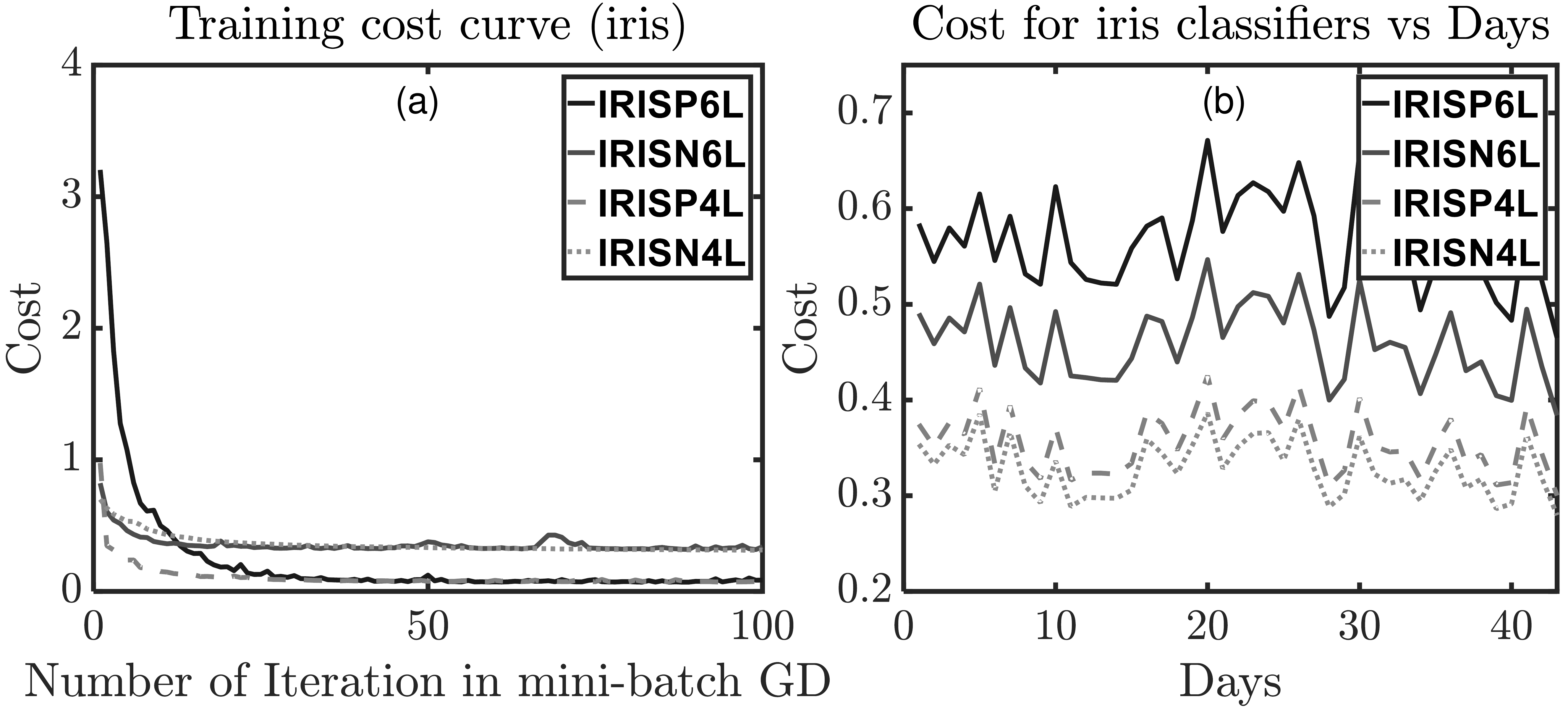}
 \end{center}
 \vspace{-1em}
 \caption{Iris classifier: (a) cost curve over the entire dataset during training in $app02$ (IRISP4L,IRISP6L) and $app03$ (IRISN4L,IRISN6L), (b) cost over 43 set of values of the qubit quality metrics of IBMQX4.} \label{fig:iris_trc}
 \vspace{-1em}
\end{figure}

\begin{figure} [] 
 \begin{center}
    \includegraphics[width=0.45\textwidth]{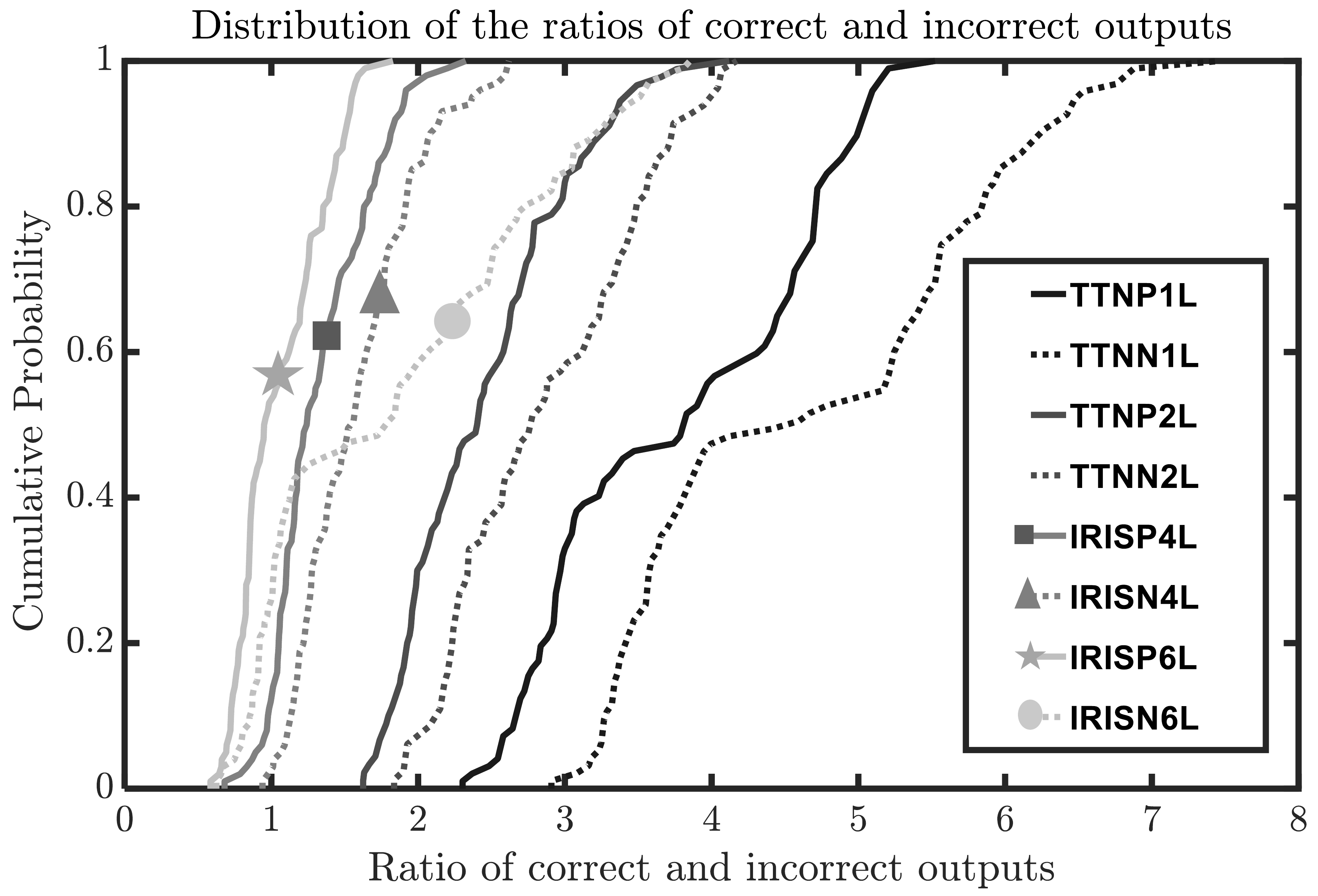}
 \end{center}
 \vspace{-1em}
 \caption{Cumulative density function of the observed ratio's between the correct and incorrect outputs for trained $PQC$'s ($app02$ and $app03$) on IBMQX4 (100 observations per $PQC$ with randomly chosen inputs and 1024 shots per observation).} \label{fig:cdf}
 \vspace{-1em}
\end{figure}

The training is done based on the $app02$ (IRISP4L, IRISP6L) and the $app03$ (IRISN4L, IRISN6L) approach with mini-batch gradient descent optimization scheme (batch size = 5 in Equation \ref{eqn:classical}). The cost curves during the training (100 iterations) are shown in Figure \ref{fig:iris_trc}(a). In Figure \ref{fig:iris_trc}(b), we have shown the actual cost (calculated using our noisy hardware simulation framework) over the entire data-set for 43 days of data of the qubit quality metrics of IBMQX4. The actual cost has been consistently smaller for the $app03$ approach as evident from Figure \ref{fig:iris_trc}(b). The average cost over the 43 sets of data for IRISP6L ($app02$) has been found to be 0.56 which is 21.7\% larger than the average cost (0.46) for IRISN6L ($app03$). 
We have executed the trained $PQC$'s on IBMQX4 hardware for the entire iris dataset and the resulting cumulative probability of the ratio's of the correct and incorrect outputs are shown in Figure \ref{fig:cdf}. The ratio values for $app03$ (IRISN4L,IRISN6L) are considerably higher than $app02$ (IRISP4L,IRISP6L) for similar values of the cumulative probability. The average value of $r$ for all the $app03$ $PQC$'s was found to be 42.5\% higher than $app02$ for the iris classifiers.

\section{Conclusions} \label{sec:end}

We presented the shortcomings of current training approaches for parameterized quantum circuits ($PQC$) and proposed a fully classical training methodology for target $NISQ$ hardware to address the impact of temporal variations in qubit quality metrics. We present a simulation framework to model the circuit behavior on a target noisy quantum hardware. We validate our proposed solutions through comprehensive simulations and experiments on a real quantum device (IBMQX4) of two quantum classifiers built with $PQC$. The proposed methodology can improve the performance of any $PQC$ based quantum application on a target $NISQ$ hardware.

\bibliographystyle{IEEEtran}

\end{document}